\documentclass[aps,prb,twocolumn]{revtex4-1}
\usepackage{graphicx}
\usepackage{amsmath}
\usepackage{amsfonts}
\usepackage{color}
\graphicspath{ {./images} }

\begin{document}

\title{Machine learning non-Hermitian topological phases}

\author{Brajesh Narayan }
\affiliation{School of Physics, University College Dublin, Belfield, Dublin 4, Ireland}
\author{Awadhesh Narayan }
\email{awadhesh@iisc.ac.in}
\affiliation{Solid State and Structural Chemistry Unit, Indian Institute of Science, Bangalore 560012, India}

\date{\today}

\begin{abstract}
Non-Hermitian topological phases have gained widespread interest due to their unconventional properties, which have no Hermitian counterparts. In this work, we propose to use machine learning to identify and predict non-Hermitian topological phases, based on their winding number. We consider two examples -- non-Hermitian Su-Schrieffer-Heeger model and its generalized version in one dimension and non-Hermitian nodal line semimetal in three dimensions -- to demonstrate the use of neural networks to accurately characterize the topological phases. We show that for the one dimensional model, a fully connected neural network gives an accuracy greater than 99.9\%, and is robust to the introduction of disorder. For the three dimensional model, we find that a convolutional neural network accurately predicts the different topological phases.
\end{abstract}

\maketitle

\section{Introduction}

The Hermitian nature of the Hamiltonian is a central postulate of quantum mechanics~\cite{dirac1981principles}. However, the investigation of systems with departure from Hermiticity has a long history~\cite{hatano1996localization,bender1998real,heiss2004exceptional,berry2004physics,bender2007making,rudner2009topological,moiseyev2011non}. Study of such open systems has been widely applied in nuclear reactions, quantum optics, photonics and mesoscopic systems~\cite{ashida2020non}.

The interest in non-Hermitian systems has seen a resurgence with a vibrant interaction with the field of topological phases -- this has resulted in a rapid flurry of activity on non-Hermitian topological phases~\cite{lee2016anomalous,gong2018topological,shen2018topological,leykam2017edge,yao2018edge,yao2018non,lee2019topological}. These exhibit remarkable properties with no counterparts in Hermitian systems, such as exceptional points~\cite{kawabata2019classification}, non-Hermitian skin effects~\cite{alvarez2018non,song2019non,lee2019anatomy,okuma2020topological} and breakdown of bulk-boundary correspondence~\cite{esaki2011edge,xiong2018does,kunst2018biorthogonal,wang2019non,borgnia2020non}, to name just a few. In addition to the rapid advancements in the theory of non-Hermitian topological systems, there have been several exciting developments in their experimental study. Photonic crystals~\cite{zhou2018observation,pan2018photonic,zhao2019non,xiao2020non,weidemann2020topological}, optical systems~\cite{zeuner2015observation,cerjan2019experimental} and topoloelectrical circuits~\cite{helbig2020generalized} have been demonstrated to be versatile platforms to investigate non-Hermitian topological phases.

In recent years, machine learning techniques have been applied, with success, to a number of physical settings~\cite{carleo2019machine,butler2018machine}. In particular, the study of different phases and phase transitions has been actively pursued in the last few years using machine learning methods~\cite{carrasquilla2017machine,wang2016discovering,wetzel2017unsupervised,zhang2017quantum,van2017learning}. Excitingly, these techniques have also been employed in identification and characterization of Hermitian topological phases of matter. These topological phases are novel phases of matter, which can not be classified by conventional Landau-Ginzburg symmetry breaking paradigm~\cite{qi2011topological,hasan2010colloquium,chiu2016classification}. Neural networks have been successfully used to learn topological invariants~\cite{deng2017machine,zhang2018machine,sun2018deep}. Unsupervised machine learning has been demonstrated to be useful for identifying topological phases~\cite{rodriguez2019identifying,scheurer2020unsupervised}. Furthermore, real space formulations of the topological invariants have been studied using artificial neural networks~\cite{carvalho2018real,holanda2020machine}. Recently, new insights into machine learning of topological quantum phase transitions have been gained~\cite{zhang2020interpreting}.

In this contribution, we introduce machine learning for non-Hermitian topological phases. Using two different examples -- non-Hermitian Su-Schrieffer-Heeger model in one dimension and non-Hermitian nodal line semimetal in three dimensions -- we demonstrate that machine learning can be used for identifying non-Hermitian topological phases based on their winding number. We discover that for the one dimensional case, a fully connected neural network yields an excellent prediction accuracy of greater than 99.9\%. We show that these predictions are robust upon introducing noise to the training data. On the other hand, for the three dimensional example, we find that the overall accuracy for the fully connected network is less than 50\%. We demonstrate that use of a convolutional neural network gives an excellent performance for this higher dimensional case, yielding an accuracy exceeding 99.8\%.

\begin{figure*}
\centering
  \includegraphics[width=6in]{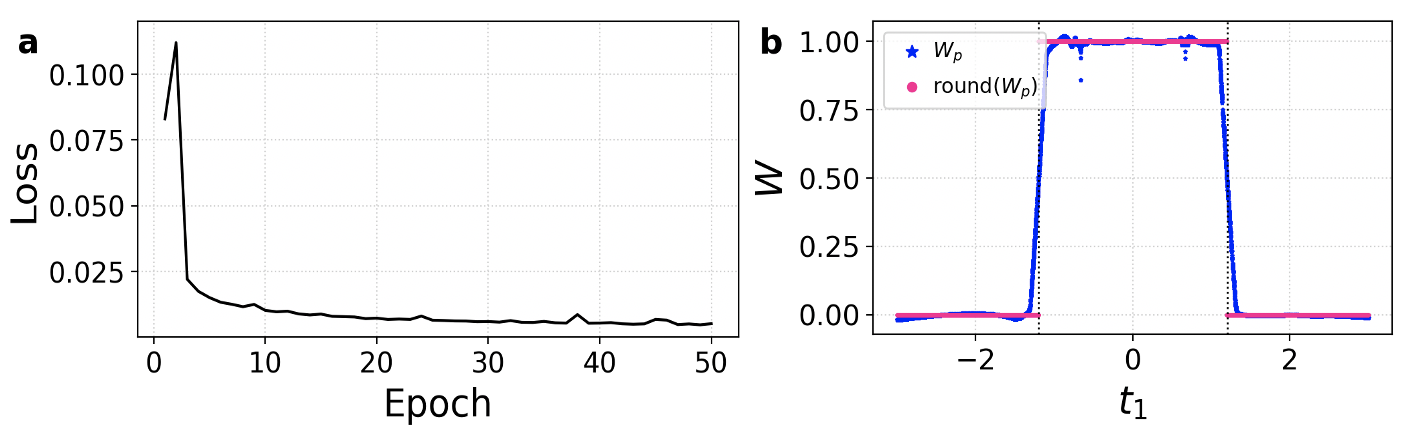}
  \caption{\textbf{Fully connected neural network for the non-Hermitian Su-Schrieffer-Heeger model.} A fully connected neural network with 2 hidden layers was constructed. The hidden layers comprised of 100 and 32 neurons, respectively. The training set consisted of $10^5$ samples. (a) The loss or cost with each epoch of training. The network was tested on a set of $10^4$ samples, not seen by the network during the training. (b) The predicted winding number, $W_{p}$, (in blue) and its rounded off value (in pink) for the test set. The network was able to predict with an accuracy of 99.98\%. The dashed vertical lines show the analytical value of $t_1$ at the phase transition. Here we have chosen $\gamma=4/3$ and $t_2=1$.
 }\label{ssh_neural}
\end{figure*}

\section{Su-Schrieffer-Heeger model} 

We begin our analysis by considering the Su-Schrieffer-Heeger (SSH) model -- the paradigmatic non-Hermitian model exhibiting topological phases~\cite{yao2018edge,lieu2018topological}. The Hamiltonian reads

\begin{equation}
    H(k)=(t_1+t_2\cos k)\sigma_x +(t_2\sin k +i\gamma/2)\sigma_y,
\end{equation}

where $\sigma_i$ ($i=x,y,z$) are the Pauli matrices and $k$ denotes the momentum. Here $t_1$ and $t_2$ are hopping strengths and a finite $\gamma$ introduces a non-Hermiticity to the Hamiltonian. The non-Hermitian topological phase of this model is characterized by the winding number, $W=1$, while the trivial phase has $W=0$~\cite{yao2018edge}. The model features several interesting aspects including the non-Hermitian skin effect as well as a breakdown of the bulk-boundary correspondence.  

We rewrite our Hamiltonian in the form $H(k)=h_x(k)\sigma_x+h_y(k)\sigma_y$ and use it as an input for our neural network at $P$ different points. Here $h_x=t_1+t_2\cos k$, $h_y=t_2\sin k +i\gamma/2$ and $k= 2\pi n/P$ ($n=0,...,P$). The input data can also be written as $(P+1)\times2$ matrices of the form

\begin{equation}
\begin{pmatrix}
h_x(0) &h_x(2\pi/P)& ... & h_x(2\pi)\\
h_y(0) &h_y(2\pi/P)& ... & h_y(2\pi)
\end{pmatrix}^T.
\end{equation}\\

The winding number, $W$, is defined as 

\begin{equation}
W= -(i/2\pi)\oint_0^{2\pi} U^*(k)\partial_kU(k)dk, 
\end{equation}

where $U(k)=h_x(k)+ih_y(k)$. For discretized data, the above winding number equation can be rewritten as

\begin{equation}
W=(1/2\pi)\sum^{P}_{n=1}\Delta\Theta(n),
\end{equation}

with $\Delta\Theta(n)=[\Theta(n)-\Theta(n-1)]$ mod $2\pi$ and $\Theta(n) = \mathrm{arg} [U(2\pi n/P)]$.

With this input, we constructed a fully connected, i.e. dense, neural network with two hidden layers. We used 100 neurons in the first hidden layer and 32 neurons for the second hidden layer. Rectified linear unit (ReLU) activation function was used for the hidden layers. To train our neural network we generated a training set with $10^5$ samples. For generating the training set, we set $t_2=1$, $\gamma=4/3$ and $t_1$ was chosen randomly from the range $[-3,3]$. The network was trained with 2000 batches, with each batch having a size of 50. The training was performed 50 times, i.e. number of epochs is 50. The loss (or cost) with each epoch of training is shown in Fig.~\ref{ssh_neural} (a). We note that the neural network converges rapidly. We checked the accuracy with each training cycle for both training and test sets and found that their accuracy are approximately equal, which rules out over-fitting.

After having trained the neural network on the training set, we use the network to predict the winding number on a test set which consisted of $10^4$ samples not seen by the network during the training. The predicted winding number, $W_p$, is presented in Fig.~\ref{ssh_neural} (b). We note that our trained neural networks yield winding numbers close to integer values and we also plot the output rounded off to the nearest integer, as is common practice~\cite{zhang2018machine}. We find that our trained neural networks show a very high accuracy of more than 99.9\%. In particular, it is able to correctly predict the values of $t_1$ at which the topological phase transition from $W=1$ to $W=0$ takes place.

\begin{figure}
\centering
  \includegraphics[width=8cm]{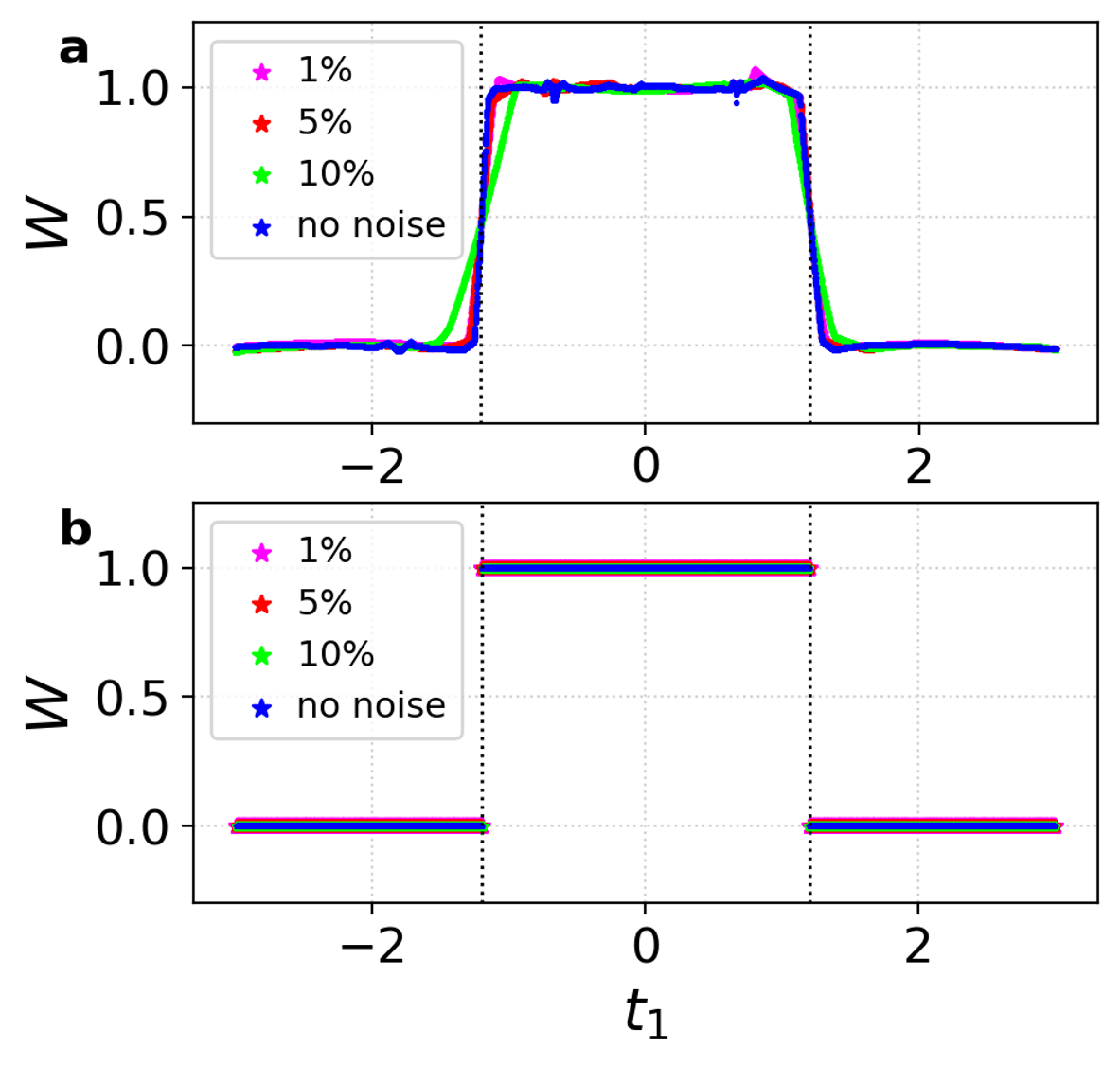}
  \caption{\textbf{Effect of noise on the predicted winding number.} Randomly sampled Hamiltonian from a computation lacks noise. Data collected from experiments would invariably show noise. To simulate this, noise was artificially introduced in the training data. This was achieved by randomly adding a value between [-0.5$t_2$,0.5$t_2$] to $h_x$ or $h_y$. The robustness of the network was checked by training the network on training sets with 1\%, 5\% and 10\% noise and testing it on a separate test set. Predicted winding numbers and their rounded off value, for training with 1\% (in magenta), 5\% (in red) and 10\% (in green) noise, are shown in (a) and (b), respectively. We obtained very high accuracy of 99.92\%, 99.90\% and 99.70\%, respectively The vertical dashed lines are the analytically obtained values of $t_1$ at the phase transition. 
}\label{ssh_noise}
\end{figure}

Our randomly sampled Hamiltonian does not include any noise. On the other hand, data collected from experiments would invariably show some degree of noise. To simulate this scenario, we artificially incorporated noise in our training data. To do so, we added a value between [-0.5$t_2$,0.5$t_2$] to $h_x$ or $h_y$. We checked the robustness of the network by training the network on training sets with 1\%, 5\% and 10\% noise and testing it on a separate test set. The resulting predictions for the winding number and their rounded off value are shown in Fig.~\ref{ssh_noise}. Remarkably, the trained neural network is very robust and we obtain a very high accuracy greater than 99.7\% in all these cases. This suggests that our neural network approach could be reliable even in the presence of noise in the input training data. We note that disorder which breaks translational symmetry is not described by the momentum space picture and consequently by the winding number -- our method is not directly applicable to these cases.

\begin{figure}
\centering
  \includegraphics[width=8cm]{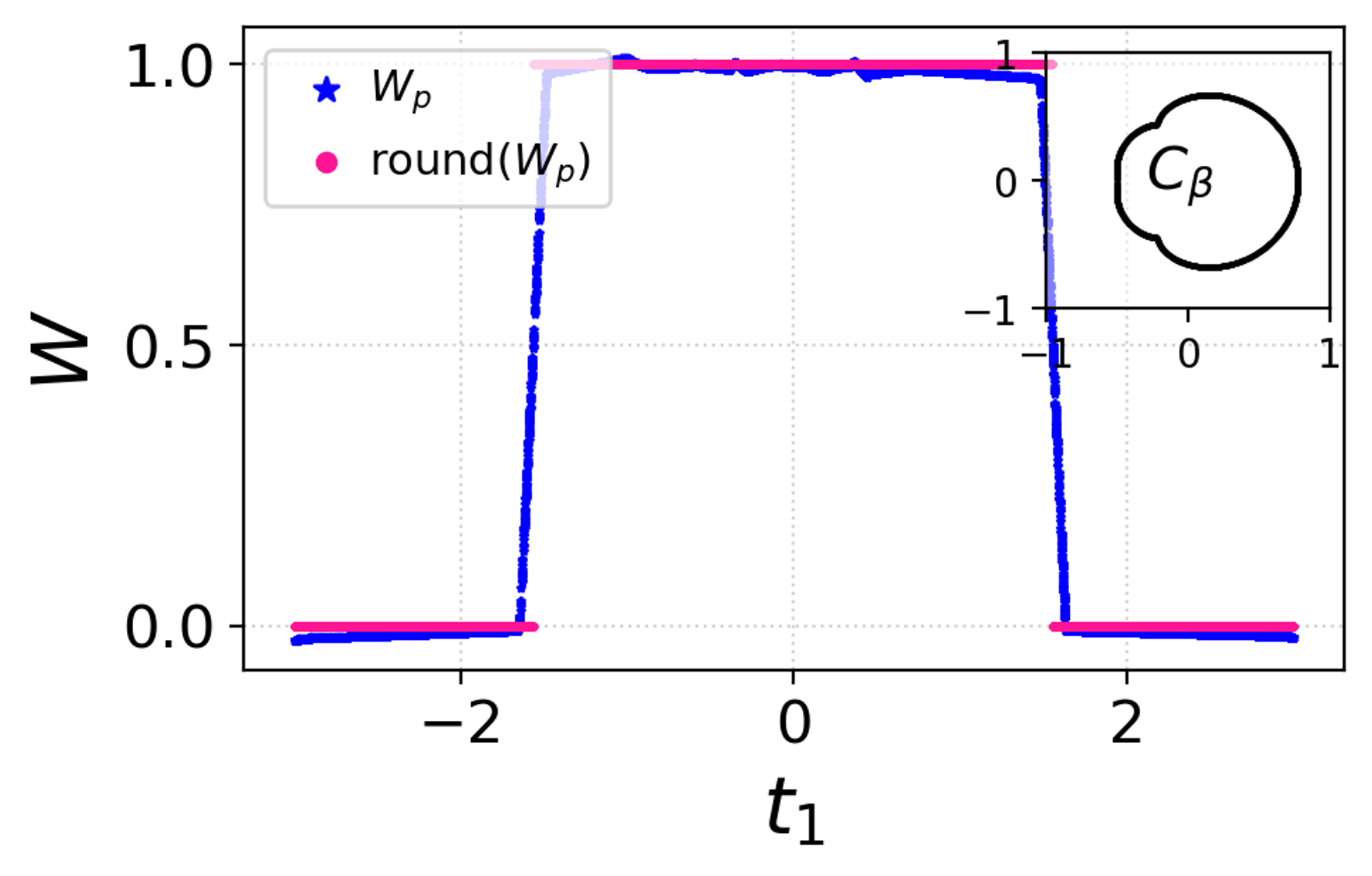}
  \caption{\textbf{Learning the non-Bloch winding number for generalized SSH model.} The predicted winding number (in blue) and the rounded off value (in pink) for the generalized non-Hermitian SSH model. A total of $10^5$ training samples were generated with $t_1\in[-3,3]$, $t_2=1$, $t_3=1/5$, $\gamma_1=4/3$, and $\gamma_2=0$. The network was tested on $10^4$ samples, not seen by the network during training. An accuracy of 99.8\% was obtained. The inset shows the generalized Brillouin zone $C_\beta$ for $t_1=1.1$. 
}\label{generalized_model}
\end{figure}

\section{Non-Bloch winding number} 

Next, we consider the generalized non-Hermitian SSH Hamiltonian~\cite{yokomizo2019non}

\begin{eqnarray}
    H(k)&=&[t_1+(t_2+t_3)\cos k+i(\gamma_2/2)\sin k]\sigma_x \nonumber \\ 
    &+&[(t_2-t_3)\sin k +i\gamma_1/2 -i(\gamma_2/2)\cos k]\sigma_y.
\end{eqnarray}

This reduces to our earlier model for $t_3=\gamma_2=0$ and $\gamma_1=\gamma$. The generalized Bloch Hamiltonian is obtained by the replacement $e^{ik}\rightarrow \beta$. Expressed in terms of $\beta$, the Hamiltonian becomes $H(\beta)=R_+\sigma_++R_-\sigma_-$, where $\sigma_{\pm}=(\sigma_x\pm i\sigma_y)/2$ and $R_{\pm}$ read

\begin{eqnarray}
    R_+&=&(t_2-\gamma_2/2)\beta^{-1}+(t_1+\gamma_1/2)+t_3\beta \nonumber,\\
    R_-&=&t_3\beta^{-1}+(t_1-\gamma_1/2)+(t_2+\gamma_2/2)\beta.
\end{eqnarray}

The eigenvalue equation is $R_+R_-=E^2$, which has four solutions, $\beta_{i}$ ($i=1,2,3,4$), in general. The trajectory satisfying the condition $|\beta_2|=|\beta_3|$ traces the generalized Brillouin zone labeled by $C_\beta$~\cite{yao2018edge,yokomizo2019non}. The non-Bloch winding number is then defined as

\begin{equation}
W=-\frac{w_+-w_-}{2},
\end{equation}

where $w_{\pm}=[\mathrm{arg}R_{\pm}]_{C_\beta}/2\pi$. Here $[\mathrm{arg}R_{\pm}]_{C_\beta}$ denotes the change of phase of $R_{\pm}$ as $\beta$ traverses $C_\beta$~\cite{yokomizo2019non}. Next, we discretize these quantities over points on $C_{\beta}$, and construct a fully connected neural network. We trained the network using $10^5$ samples, with $t_1\in[-3,3]$, $t_2=1$, $t_3=1/5$, $\gamma_1=4/3$, and $\gamma_2=0$. We next test the network on $10^4$ samples, which were previously not seen by the network, and calculate the non-Bloch winding number over the $C_\beta$. The results are shown in Fig.~\ref{generalized_model}. Our trained neural network performs very well, yielding an accuracy of 99.8\%. This shows that our formalism is successful in learning this non-Bloch winding number and is thus applicable in a more general setting.

\section{Nodal line semimetal model} 

We now consider a higher dimensional model to understand if the network is able to learn the winding number and phase transitions in a more general setting. The non-Hermitian continuum model for a nodal line semimetal reads~\cite{wang2019non}

\begin{equation}
H=(m - Bk^2)\sigma_x + (v_zk_z + i\gamma_z)\sigma_z,      
\end{equation}

where $k=\sqrt{k_{x}^2+k_{y}^2+k_{z}^2}$ and $v_z$ is the Fermi velocity. The parameters $m$ and $B$ control the existence and radius of the nodal line in the Hermitian limit. This model shows a rich phase diagram with winding numbers 0,-1/2 and -1, in addition to exceptional rings and the non-Hermitian skin effect~\cite{wang2019non}. For this more general case, the input Hamiltonian is  $H(k_z) = h_x(k_z)\sigma_x+h_z(k_z)\sigma_z$, where $h_x=m-B(k_{x}^2+k_{y}^2+k_{z}^2)$ and $h_z=v_zk_z+ i\gamma_z$. We treat $k_x$ and $k_y$ as parameters and discretize $k_z=2\pi n/P$ where $n\in \mathbb{Z}$, such that $n=-P, -P+1, ... ,P-1, P$ and $n \neq 0$. The input data can be expressed as $2P\times 2$ matrices. Analogous to the Su-Schrieffer-Heeger model, the winding number for discrete data can be computed using $W=(1/2\pi)\sum_{n}\Delta\Theta(n)$. Here $\Delta\Theta(n)=[\Theta(n)-\Theta(n-1)]$ mod $2\pi$ and $\Theta(n) = \arctan (h_x/h_z)$. Using this input data, we first constructed a fully connected neural network. Our training set consisted of $ 8\times10^5$ Hamiltonians with $k_x$ and $k_y$ uniformly distributed in the range $[-1,1]$. The network was subsequently used to predict winding numbers on a test set which consisted of $ 2\times10^5$ Hamiltonians in the same range of $k_x$ and  $k_y$. Test samples were not seen by the network during the training. Surprisingly, we found that the overall accuracy of our fully connected network for this higher dimensional model is less than 50\%, which is no better than a random guess. We were unable to improve the accuracy of the predictions of the fully connected network by changing its architecture.

\begin{figure}
\centering
  \includegraphics[width=8cm]{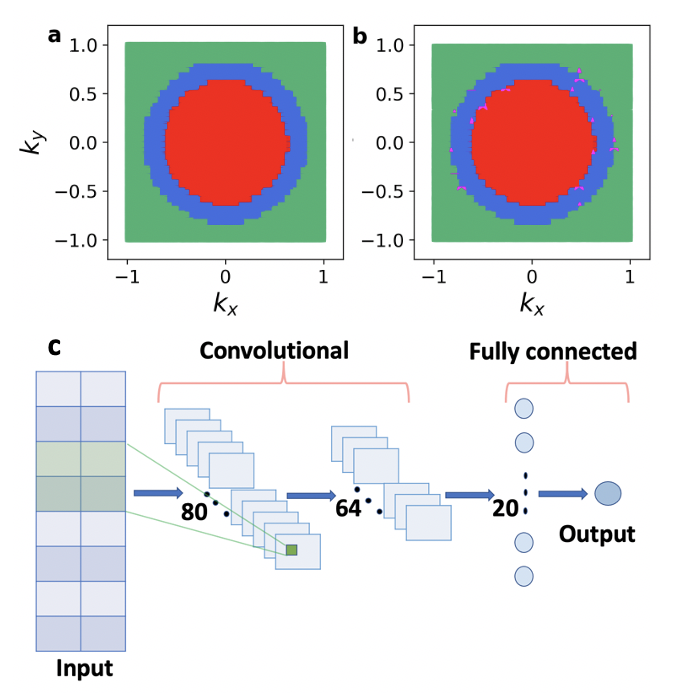}
  \caption{ \textbf{Convolutional neural network for non-Hermitian nodal line semimetal model.} (a) Computed and (b) predicted winding number as a function of $k_{x}$ and $k_y$, when $m=0.4$, $\gamma_z=0.2$, $v_z=B=1$ and $k_z=0$. Areas with winding numbers 0, -1/2 and -1 are shown in green, blue and red, respectively. We obtained an accuracy of 99.95\%. In (b) the incorrect predictions are marked in magenta. These occur predominantly near the phase boundaries. (c) Schematic of our convolutional neural network with 2 convolutional layers, with 80 and 64 filters and kernel size of $2\times2$ and $1\times1$, followed by a fully connected layer with 20 neurons before the output layer, which was used to predict the winding numbers.
 }\label{nlsm_neural}
\end{figure}

To overcome this limitation, we next constructed a more sophisticated convolutional neural network with 2 convolutional layers, each comprising of 80 and 64 filters with a kernel size of $2\times2$ and $1\times1$, followed by a fully connected layer with 20 neurons before the output layer [see Fig.~\ref{nlsm_neural} (c)]. We used a similar training as in the case of the fully connected network and employed the trained network to predict the winding numbers on a test set. Our results are presented in Fig.~\ref{nlsm_neural}, where a comparison between the calculated [panel (a)] and predicted [panel (b)] phase diagrams is shown. The green, blue and red regions in the $k_x-k_y$ plane correspond to winding numbers of 0, -1/2 and -1, respectively. The two plots bear a remarkable resemblance and our overall prediction accuracy is 99.95\%, using only 50 training cycles. We also notice a few tiny patches of incorrect predictions (shown in magenta), which occur predominantly near the phase boundaries between regions with different winding numbers. Overall, our convolutional neural network is reliable and suitable for predictions involving higher dimensions and several topological phases.

To gain more insight into our convolutional neural network, whose schematic is shown in Fig.~\ref{nlsm_neural} (c), we investigate the details of its learning. Our convolutional neural network consists of three hidden layers, two convolutional layers and one fully connected layer. The first layer in the network is a convolutional layer with 80 filters, hence 80 different convolutions are performed with respect to the input Hamiltonian. 

\begin{equation}
\begin{split}
B^i(n) = &f(A^i_{11}h_x(2\pi(n-1)/P) + A^i_{12}h_z(2\pi(n-1)/P) \\ 
&+ A^i_{21}h_x(2\pi n/P) + A^i_{22}h_z(2\pi n/P) + A^i_0,   
\end{split}
\end{equation}

where $A^i_{\alpha\beta}$ is a $2\times2$ kernel, $i=1,...,80$, $n\in \mathbb{Z}$ \& $n= [-P+1,0) \cup (0,P]$, $\alpha,\beta=1,2$ and $f(x)$ is the activation function. The second layer is also a convolutional layer with 64 filters. Here the convolutions are performed using a $1\times1$ kernel, $C^i$. The output of this layer, 

\begin{equation}
D^i(n)=f\left(\sum^N_{i=1}C^iB^i(n)+ C^i_0\right),
\end{equation}

is equivalent to the $\Delta\Theta (n)$ of the winding number formula. In the third and the final hidden layer, which is fully connected, the network attempts to add all $\Delta\Theta (n)$ to output the winding number. In the final layer, all the 20 neurons of the last hidden layer are mapped on to a single output neuron to yield the predicted winding number

\begin{equation}
W_p=\sum^{20}_{q=1}F_qE_q+ G.
\end{equation}

In the above, $E_n=f\left(\sum^N_{i=1}M_{qn}D(n)+ N_q\right)$ with $q=1,...,20$. The network successfully determines all the fitting parameters, $A^i$, $C^i$, $M_{qn}$, $N_q$, $F_q$ and $G$, during the training. With these insights, we can conclude that the network is capable of learning the winding number formula in cases with co-existence of several different topological phases. This is a reliable and efficient approach to characterize non-Hermitian topological phases and the understanding gained from the scrutiny of its inner workings would be useful for formulating extensions to other systems. It would further be interesting to explore the application of machine learning to other intriguing properties of non-Hermitian systems, such as exceptional points and non-Hermitian skin effects.

\section{Summary and outlook} 

We demonstrated the use of machine learning to identify non-Hermitian topological phases, characterized by their winding numbers. For the one-dimensional non-Hermitian SSH model, we trained a fully connected neural network to predict the different phases with an accuracy greater than 99.9\%. For a three-dimensional non-Hermitian nodal line semimetal model, we constructed and trained a convolutional network to yield excellent accuracy in predictions of the topological phases. Our proposed methods could be potentially useful for machine learning of other non-Hermitian topological phases~\cite{cerjan2018effects,yang2019non,rui2019pt,rui2019topology,banerjee2020non,carlstrom2019knotted,yang2019non1,yang2020jones}, and could be generalized to include those with disorder~\cite{tang2020topological,claes2020skin}. Furthermore, we envisage that our methods could be applied for identification of non-Hermitian topological phases in future experiments. \\

\noindent \textit{Note added--} After completion of this work we became aware of complementary studies in Refs.~\onlinecite{zhang2020machine,yu2020unsupervised}.

\section*{Acknowledgments} 

BN acknowledges financial support received from Thomas Preston PhD Scholarship. AN acknowledges support from the start-up grant (SG/MHRD-19-0001) of the Indian Institute of Science and DST-SERB (project number SRG/2020/000153).

%\bibliography{references}


\end{document}